\documentstyle[aps]{revtex}
\input epsf.tex
\def \Laa{$L_{a}$}

\def \beq{\begin{equation}}
\def \eeq{\end{equation}}
\begin{document}
\baselineskip=24pt
\begin{center}
\bf{Mean field analysis of the
$SO(3)$ lattice gauge theory at finite temperature } \\
%\vspace{0.2cm}

\vspace{1cm}

\rm{Srinath Cheluvaraja}\footnote[1]{e-mail:srinath@theory.tifr.res.in} \\
Tata Institute of Fundamental Research \\
Mumbai 400 005,
India
\end{center}
\noindent{\bf{ABSTRACT}}\\

We study the finite temperature properties of the 
$SO(3)$ lattice gauge theory using mean field theory.
The main result is the calculation of the effective action at
finite temperature. 
The form of the effective action is used to explain the behaviour
of the  
adjoint Wilson line in numerical
simulations.
Numerical simulations of the $SO(3)$ lattice gauge theory
show that the adjoint Wilson line has
a very small value at low temperatures; at high temperatures, metastable states
are observed in which the adjoint Wilson line takes positive or negative
values.
The effective action is able to explain
the origin of these metastable states. 
A comparison of the effective actions of the $SU(2)$ and the $SO(3)$
lattice gauge theories
explains their different behaviour at high
temperatures. The mean field theory also predicts a finite temperature
phase transition in the $SO(3)$ lattice gauge theory.

\vspace{0.5cm}
\begin{flushleft}
\end{flushleft}

\newpage
\begin{section}{Introduction}
Confining gauge theories
are expected to pass over into a deconfining phase at high
temperatures. The first explicit non-perturbative calculation \cite{suss}
to show this was done in the strong coupling limit of lattice gauge theories
(LGTs).
Since then, there have been many studies of the finite temperature
properties of LGTs.
It is hoped that an understanding of their 
properties will shed some light on the high temperature phase
of Yang-Mills theories. There have been numerous studies of the
finite temperature properties of $SU(2)$ \cite{early,yaffe,su2is} and $SU(3)$ 
\cite{mont} LGTs.
The basic observable that is studied in these systems 
is the Wilson-Polyakov line
(henceforth called the Wilson line) which is defined as
\beq
L_{f}(x)= Tr_{f} \exp i\int_{0}^{\beta}  A(x,x_{4})dx_{4} \quad .
\eeq
The subscript $f$ indicates that the
trace is taken in the fundamental representation of the
group. 
The Wilson line  has the physical interpretation of measuring the free
energy ($F(x)$) of a static quark in a heat bath at a temperature $\beta^{-1}$.
This is made explicit by writing it in the form:
\beq
\langle L_{f}(x) \rangle = \exp (-\beta F(x)) \quad ;
\eeq
a non-zero value of the Wilson line implies that a
static quark has a finite free energy whereas a zero-value implies
that it has infinite free energy.
The strong coupling analysis in \cite{suss} shows that the
Wilson line remains zero at low temperatures and becomes non-zero at
high temperatures \cite{suss}, signalling a finite temperature 
confinement to deconfinement phase
transition.
This transition is also observed in numerical simulations.
The action for the $SU(2)$ LGT is usually taken to be the Wilson
action \cite{wils} and is given by
\beq
S=(\beta_{f}/2)\sum_{n\ \mu \nu}tr_{f}\ U(n\ \mu \nu)\quad ;
\eeq
the subscript $f$ indicates that the
trace is taken in the fundamental representation of $SU(2)$.
The variables $U(n\ \mu \nu)$ are the usual 
oriented plaquette variables:
\beq
U(n\ \mu \nu)=U(n\ \mu)U(n+\mu\ \nu)U^{\dagger}(n+\nu\ \mu)U^{\dagger}(n\ \nu)
\quad ;
\eeq
the $U(n\ \mu)$s are the link variables which are elements of the group $SU(2)$.
A finite temperature system (at a temperature $\beta^{-1}$) 
is set up by imposing periodic
boundary conditions (with period $\beta$) in the Euclidean time direction.
This results in an additional global $Z(2)$ symmetry that acts on the 
temporal
link variables as follows:
\beq
U(n\ n_{4})\rightarrow Z U(n\ n_{4}) \quad .
\eeq
$Z$  is an element of the center of the group $SU(2)$ and
takes the values $+1$ or $-1$.
Under the action of this symmetry transformation, the Wilson line transforms as
\beq
L(x)\rightarrow Z L(x) \quad .
\eeq
It is evident that the high temperature phase
(in which the Wilson line has a non-zero average value) breaks this global symmetry. 
As a result of this symmetry, the high temperature
phase of the $SU(2)$ LGT is doubly degenerate and the two states are
related by a $Z$ transformation. 
The two degenerate states have the same
free energy because of this global symmetry.
Numerical simulations observe these states as metastable states in which
the Wilson line takes two different values which are related by a $Z$
transformation.
The role of the center symmetry was further emphasized in \cite{yaffe}
where it was argued that the order of the transition to the high
temperature phase could be understood in terms of the universality
classes present in 3-d spin models having this symmetry. These expectations
have been borne out for the $SU(2)$ \cite{su2is} and 
the $SU(3)$ \cite{mont} LGTs in which
one observes a second order Ising like and a first order $Z(3)$ like 
phase transition respectively.

Another choice of an action, which is expected to 
lead to the same physics as the
Wilson action, is the adjoint action that is given by
\beq
S=(\beta_{a}/3)\sum_{n\ \mu \nu}tr_{a}\ U(n\ \mu \nu) \quad .
\eeq
Here, the subscript $a$ denotes that the trace is taken in the
adjoint representation of $SU(2)$. The trace in the adjoint representation
can be expressed in terms of the trace in the fundamental representation
as
\beq
tr_{a}U=tr_{f}U^2-1 \quad .
\eeq
From its definition, the adjoint action describes an $SO(3)$ LGT
since the link variables $U(n\ \mu)$ and $-U(n\ \mu)$ have the same
weight in the action. 
Unlike the $SU(2)$ LGT, the $SO(3)$ LGT has a bulk (zero temperature)
transition at $\beta \approx 2.5$. This transition is understood in terms
of the decondensation of $Z(2)$ monopoles \cite{hall}. An interesting 
and important question is
whether the $SO(3)$ LGT has a deconfinement transition like the
$SU(2)$ LGT. The universality of lattice gauge theory actions would require
$SU(2)$ and $SO(3)$ LGTs to have the same continuum limit.
We will show that our mean field analysis does predict
a deconfinement transition for the $SO(3)$ theory.
In the $SO(3)$ theory, the appropriate observable 
( though it is not an
order parameter in the strict sense)
to study deconfinement
is the Wilson line in the adjoint representation \cite{our}; this 
observable is defined as
\beq
L_{a}(x)= Tr_{a} \exp i\int_{0}^{\beta}  A(x,x_{4})dx_{4} \quad .
\eeq
The subscript $a$ denotes the trace in the adjoint representation.
The Wilson line in the fundamental representation is always zero in this
model because of a local $Z$ symmetry. This will be explicitly shown later.
The adjoint Wilson line can also be interpreted as measuring the free energy of
a static quark in the adjoint representation by writing it as
\beq
\langle L_{a}(x) \rangle=\exp (-F_{a}(x)) \quad .
\label{free}
\eeq
The $Z$ symmetry acts trivially on this observable.
A further generalization of the Wilson action
is the mixed action LGT \cite{creu} that is defined as
\beq
S=(\beta_{f}/2)\sum_{n\ \mu \nu}tr_{f}\ U(n\ \mu \nu)+
(\beta_{a}/3)\sum_{n\ \mu \nu}tr_{a}\ U(n\ \mu \nu) \quad .
\eeq
The finite temperature properties of this model have been studied
in \cite{gav}.

The two Wilson lines
can be expressed as a function of the gauge invariant variable $\theta$ as
\beq
L_{f}(x)=2 \cos(\theta/2)\ \ 
L_{a}(x)=1+2 \cos(\theta) \quad ;
\eeq
$\theta$ is the phase of the eigenvalues of
\beq
P \exp i\int_{0}^{\beta}A(x,x_{4})dx_{4} \quad.
\eeq
The variable $\theta$ is gauge invariant and can be used to
characterize the various phases of the system.

It is the purpose of the present note to find the effective action,
$V_{eff}(\theta)$, for the $SO(3)$ LGT at non-zero temperatures. The
effective action is calculated in the mean field approximation. The effect of
the fluctuations about the mean field solution 
is also considered and they are shown to be quite important at low
temperatures.
The effective action is also calculated for the $SU(2)$ LGT and the
differences are pointed out with the $SO(3)$ LGT. We then make some comments
on the mixed action LGT.
Though a mean field analysis of the $SO(3)$ LGT is of interest in
itself,
the main motivation
for our present analysis is to qualitatively understand some of the
observations made in numerical simulations of the $SO(3)$ LGT.
Numerical studies of the $SO(3)$ LGT \cite{our} show that the adjoint
Wilson line (AWL) remains close to zero at low temperatures and
jumps to a non-zero value at high temperatures.
Both, the low and the high temperature behaviour of the adjoint Wilson line
are quite puzzling.
The small value of the
AWL at low temperatures is surprising because a static source in the adjoint 
representation
can always combine with a gluon and form a state with a finite 
free energy.
More surprising, however,
is the observation of two
distinct metastable states for the AWL at high
temperatures \cite{our}. We explain later why we call these states as
metastable states.
In numerical simulations, we find metastable states
with the AWL taking a positive or negative value depending 
on the initial configuration of the Monte-Carlo run.
A hot start ( random initial configuration)
usually settles
to a negative value whereas a cold start (ordered initial configuration) 
always settles to a positive value.
This metastability is seen even at very high temperatures.
Fig.~\ref{la} is a typical run time history of the AWL for hot and cold
starts in the high temperature phase.
The values of other observables like the plaquette
square and the $Z(2)$ monopole density (which is almost equal to zero)
are almost the same
in both these metastable states. 
All
this appears very reminiscent of the metastable states 
(of the fundamental Wilson line) observed in the high
temperature phase of the $SU(2)$ LGT in which two degenerate states related
by a $Z$ transformation are observed. Nonetheless, as there is
no obvious
symmetry in the $SO(3)$ LGT connecting the two observed metastable states
, the presence of two exactly degenerate minima
in the free energy would be quite remarkable. 
A measurement of the
correlation function of the adjoint Wilson line indicated that the
correlation lengths were the same in the $L_{a}$ positive and $L_{a}$ negative
states \cite{saum}. The
authors in \cite{saum} use this result to argue that the two states are
physically equivalent.
We will show the existence of these metastable states 
at high temperatures using mean
field theory.
The mean field analysis shows that there are minima in the effective action
at positive and negative values of $L_{a}$. The difference 
in free energy density
between these
minima depends on two parameters, $N_{\tau}$ and $\beta_{a}$;
$N_{\tau}$ is the temporal extent of the lattice 
(or the inverse temperature), and $\beta_{a}$ is the
coupling constant of the $SO(3)$ LGT.
For a range of values of the parameters, $N_{\tau}$ and $\beta_{a}$, these
minima are almost equal to each other. This may explain 
why both states are observed
in numerical simulations. 
The mean field theory analysis can be done
for the $SU(2)$ LGT as well, and the differences are pointed out with
the case of the $SO(3)$ LGT. In particular, it is shown why the state
in which the AWL takes a
negative value is absent in the $SU(2)$ theory.
Finally, we extend the mean field theory to the $SU(2)$
mixed action LGT. We conclude with a discussion of some theoretical
issues connected with the adjoint Wilson line.

The usual approach of doing a mean field theory at non-zero
temperature requires a strong coupling approximation as in
\cite{polon}. 
There are other variants of this mean field theory which
are all basically based on the idea of ignoring the 
effect of the spatial plaquettes \cite{vari}.
Spatial plaquettes tend to deconfine the system; a deconfinement
transition in the absence of spatial plaquettes will necessarily
imply such a transition with them included.
If one considers a reduced model with the spatial plaquettes
discarded,
the spatial links can be exactly integrated using a character
expansion. This leads to an effective theory of Wilson lines in
three dimensions. 
Before we present the details of this calculation, we
would like to say that there is no qualitative change in the
finite temperature properties of the system in this limit.
The spatial degrees of freedom can be considered to be inert across
the deconfinement transition, and the only role they play is to possibly
shift the transition temperature. Symmetry properties are also not in anyway
altered in this reduced model, and
even the order of the phase
transition, if there is any, should be unaffected by this simplification (this
will be shown for the $SU(2)$ theory).
In this limit of the $SO(3)$ LGT, the \Laa \  positive
and \Laa \  negative states are again observed
in numerical simulations,
just as in the full model, and they again display
the same features as in the full model.
The approximation of discarding the spatial plaquettes 
does not introduce anything extraneous into
the finite temperature properties. Even the bulk properties of the system
should remain unchanged in this approximation
because ignoring the spatial plaquettes 
gives a zero weight to the $Z(2)$ monopoles which are known to drive the
bulk transition \cite{hall} in the $SO(3)$ LGT.
In the $SO(3)$ LGT, the $Z(2)$ monopoles 
anyway donot
cost any energy because of the square term in the action. 
The main motivation for analysing the reduced
model is that an accurate mean field analysis can be
made. 

The reduced model is defined as
\beq
S=\sum_{p \in t}\chi(U) \quad;
\eeq
the summation is only over the temporal plaquettes. $\chi(U)$ is a class
function defined on the plaquette variables. We shall be concerned with
three possible forms that this function can take. They are
\beq
\chi(U)=\frac{\beta_f}{2} tr_{f}\ (U(p)) \quad ;
\eeq
this is Wilson's action for the $SU(2)$ LGT. Then we will consider
\beq
\chi(U)=\frac{\beta_a}{3} tr_{a}\ (U(p)) \quad ;
\eeq
this is the adjoint action and describes an $SO(3)$ LGT.
Finally, we will consider the mixed action,
\beq
\chi(U)=\frac{\beta_{f}}{2}\chi_{f}(U)+
\frac{\beta_{a}}{3}\chi_{a}(U) \quad .
\eeq
The character
expansion of the exponential gives
\beq
Z=\int [DU] \prod_{t}\sum_{j}\tilde \beta_{j} \chi_{j} (U(p)) \quad .
\eeq
The characters are given by the formula
\beq
\chi_{j}(\Omega)=\frac{\sin((j+1/2)\theta)}{\sin (\theta/2)} \quad .
\eeq
Here $\Omega$ denotes some $SU(2)$ group element which is parametrized
in the usual way as
\beq
\Omega=\cos(\theta/2)+i\vec \sigma .\vec n \sin(\theta/2) \quad .
\eeq
The $\tilde \beta_{j}$ can be calculated using the orthonormality
property of the characters
\beq
\int [dU] \chi_{r}(U) \chi_{s}^{*}(U)=\delta_{rs} \quad .
\eeq
The character coefficients are given by
\beq
\tilde \beta_{j}=\int [dU] \exp(S(U)) \chi_{j}^{*}(U) \quad ;
\label{chars}
\eeq
$S(U)$ can be the action for the $SU(2)$,$SO(3)$ or the mixed action
LGT.

The spatial links can be integrated using the orthogonality relation
\beq
\int [DU] D_{m_{1}\ n{1}}^{j}(U)D_{m_{2}\ n_{2}}^{k}(U^{\dagger})
=\frac{1}{2j+1}\delta_{j,k}\delta_{n_{1}\ m_{2}} \delta_{m_{1}\ n_{2}}
\quad .
\eeq
This leads to the effective 3-d spin
model with the partition function (with $\chi_{j}(\Omega (\vec r))$ acting as
the spin degree of freedom)
\beq
Z=\int [d\Omega(\vec r)]\prod_{\vec r^{\prime} \vec r}\sum_{j}(\frac{\tilde \beta_{j}}
{2j +1})^{N_{\tau}} \chi_{j} (\Omega(\vec r)) 
\chi_{j}(\Omega(\vec r^{\prime})) \quad .
\eeq
The effective action is
\beq
S_{eff}=-\sum_{\vec r \prime \vec r} \log \sum_{j} (\frac{\tilde \beta_{j}}
{2j+1})^{N_{\tau}} \chi_{j}(\Omega(\vec r))
\chi_{j}(\Omega(\vec r^{\prime})) \quad .
\label{su2}
\eeq
The partition function of this spin model can be written as
\beq
Z=\int [d\Omega(\vec r)] \exp (-S_{eff}) \quad .
\eeq
The measure is the $SU(2)$ Haar measure
\beq
d\Omega=\int_{0}^{4 \pi}\frac{d\theta (\vec r)}{4\pi}(1-\cos(\theta(\vec r)) \quad .
\eeq
So far, the analysis does not distinguish between the groups $SU(2)$ or
$SO(3)$.
The difference between them arises in the 
coefficients in the character expansion.
In the $SU(2)$ LGT, all the character coefficients are in general
non-zero and they are given by the formula:
\beq
\tilde \beta_{j}=2(2j+1)I_{2j+1}(\beta_{f})/\beta_{f} \quad .
\eeq
In the $SO(3)$ LGT, the $\tilde \beta_{j}$ are non-zero only for integer values
of $j$ and the coefficients are given by the formula:
\beq
\tilde \beta_{j}=\exp (\beta_{a}/3) (I_{j}(2 \beta_{a}/3) -I_{j+1}(2\beta_{a}/3))
\quad .
\eeq 
For the mixed action LGT, all the character coefficients are non-zero but
an expression similar to the one for $SU(2)$ and $SO(3)$ is not
available, and the chjaracter coefficients have to be determined numerically.
The properties of the character coefficients 
lead to an important difference between the effective spin models for the
$SU(2)$ and the $SO(3)$ LGTs.
Since the $SO(3)$ theory involves only the integer representations of $SU(2)$,
the following relation is true for all the spins:
\beq
\chi_{j}(\theta(\vec r) + 2\pi)=\chi_{j}(\theta (\vec r))
\quad .
\eeq
This means that the transformation
\beq
\theta (\vec r) \rightarrow \theta (\vec r) + 2 \pi \quad
\label{symm}
\eeq
is true at any single site. The above transformation is a local symmetry
of the $SO(3)$ LGT.
In the $SU(2)$ LGT, the following relation is true for the half-integer
representations:
\beq
\chi_{j}(\theta(\vec r) + 2\pi)=-\chi_{j}(\theta (\vec r))
\eeq
In the $SU(2)$ theory, the transformation in Eq.~\ref{symm}
is a symmetry only if it is 
performed simultaneously at every site. Thus the $SU(2)$ theory
has only the following global symmetry:
\beq
\theta (\vec r) \rightarrow \theta (\vec r) + 2 \pi \quad
\quad ;
\eeq
the $SO(3)$ theory has this symmetry as a local symmetry.
Under these symmetry transformations, the fundamental and adjoint Wilson
line transform as
\beq
L_{f}(\vec r)\rightarrow -L_{f}(\vec r) \\ \nonumber
L_{a}(\vec r)\rightarrow L_{a}(\vec r) \\
\quad .
\eeq
In the $SO(3)$ theory, this local symmetry (we will call it a local $Z$
symmetry)
ensures that the
expectation value of the fundamental Wilson line is always zero.

We now look for a translationally invariant solution 
that minimizes the action in this model.
This leads to the effective action:
\beq
\frac{1}{N} S_{eff}(\theta)= -\log(1-\cos(\theta))-3 
\log (\sum_{j} \frac{\tilde \beta_{j}}{2j+1} (\chi_{j}(\Omega)^2) \quad .
\eeq
The factor of three is present because we are dealing with a three dimensional
spin model. The measure term has also been absorbed in the action.
The partition function of the effective model is
\beq
Z=\int_{0}^{4\pi} [d\theta]\exp (-S_{eff}(\theta)) \quad .
\eeq
To get the effective action
we have to deal with the infinite summation over $j$. Since the higher
order terms in the character expansion are much smaller, the summation
can be terminated at some large value of $j$. This approximation
does not alter the results in any way as we have checked.
We plot the effective action for the $SU(2)$ and $SO(3)$ LGTs
as a function of $\theta$. In the plot, the range of $\theta$ is restricted to 
vary from $0$ to
$2\pi$ since the other half gives no additional information.
$\theta$ is the translationally
invariant single site value of the phase of the Wilson line; it is
a gauge invariant quantity.
The shape of the effective action depends
on the two parameters, $N_{\tau}$ and $\beta_{a}\ or\ \beta_{f}$. Depending on 
their values,
the effective action develops one or more
minima. The effective action for the $SU(2)$ theory 
for different values of $\beta_{f}$
is shown in Fig.~\ref{su2pot}.
At low temperatures, $V_{eff}(\theta)$ has the
shape of a bowl with a very broad minimum at
$\theta \approx \pi$. As the temperature increases, two minima start
developing very close to the $\theta \approx \pi$ minimum and start
receding away; at higher temperatures, these minima 
approach $\theta \approx 0$
and $\theta \approx 2\pi$.
The two minima at high temperatures are
the two states with a non-zero $L_{f}$ which differ by a $Z$
symmetry ($\theta \rightarrow 2 \pi + \theta$) and they are the 
two phases
with spontaneously broken $Z$ symmetry.
Both these states have the
same value of \Laa \ .
These two minima in the effective action represent the deconfined phase
of the $SU(2)$ LGT.
The second order nature of the
phase transition is also manifest from the evolution of the effective potential.
This second order transition is seen in
simulations of the $SU(2)$ LGT, and is also in accordance with the universality arguments in
\cite{yaffe}. We have demonstrated this result for the $SU(2)$ LGT,
even though it is a well known one \cite{polon}, simply because in our way
of doing the mean field theory we use the phase of the eigenvalues of
the Wilson line and not the trace of the Wilson line as is done in
\cite{polon}. It also serves to show that a truncation of the spatial
plaquettes does not change the finite temperature properties of the system.
We now turn to the $SO(3)$ LGT theory which is our main interest.
As we have mentioned before, the $SO(3)$ theory has a local $Z$ symmetry and this
is an important difference that we have to keep in mind.
The effective action is shown in
Fig.~\ref{so3pot}. At low temperatures, the effective action 
again develops the shape of
a bowl with a very broad minimum at $\theta \approx \pi$.
As the temperature is increased, the effective
action evolves quite differently from the $SU(2)$ theory.
The major difference from the $SU(2)$ theory is that the minimum at
$\theta \approx \pi$ always remains a minimum.
The broad minimum at $\theta \approx \pi$ gets sharper, and
minima at $\theta \approx 0,2\pi$ start developing. 
The minimum at $\theta \approx \pi$ would correspond to a value of
$L_{a}$ equal to -1 and the minima at $\theta \approx 0,2\pi$ would
correspond to a value of +3.
The minima at $\theta \approx 0,2\pi$ have
the same depth while the minimum at $\theta \approx \pi$ has a slightly
different depth.
The difference in the action between the two
states depends on the values of $N_{\tau}$ and $\beta_{a}$. For the values of the
parameters shown in the plot, the difference in depth of the 
minima at $\theta \approx 0,2\pi$ and the minimum at $\theta \approx \pi$ is small
compared to  the absolute value of these minima.
For much larger values of $\beta_{a}$, the minima at $\theta \approx 0,2\pi$
sink below the minimum at $\theta \approx \pi$. Nevertheless,
$\theta \approx \pi$ still remains a minimum, although it is 
only a local minimum. 
This evolution of the effective action signals
a phase transition at large $\beta_{a}$ across which the global minimum of the
effective action shifts from $\theta \approx \pi$ to $\theta \approx 0,2\pi$. 
Though there are two minima in the effective action at the $\theta \approx
0,2\pi$, the local symmetry ensures that the average value of the
fudamental Wilson line is always zero.
The value
of $L_{a}$ is the same at $\theta \approx 0$ and $\theta \approx 2\pi$. 
Hence, the  value of $L_{a}$ in the minima at $\theta \approx 0,2\pi$ is
the same as its value in the high temperature phase of the $SU(2)$ theory.
We can then conclude that the global minima at high temperatures in the $SO(3)$
theory correspond to a deconfining phase just as in the $SU(2)$ theory, the
only difference being that the adjoint Wilson line should be used to
label the deconfining phase. As we have mentioned before, the 
average value of the fundamental
Wilson line is zero because of the local symmetry.
The minimum at $\theta \approx \pi$ is a new feature of the $SO(3)$ theory
which is not present in the $SU(2)$ theory.
We will now compare the results of our mean field calculation
with the observations made in numerical simulations.
To make this comparison,
it is instructive to compare the distributions of the fundamental and
adjoint Wilson lines ( at a single site because the variable $\theta$ is
the phase variable at a single site) observed in numerical simulations
with the shape
of the effective action.
The distribution of $L_{a}$ in the low and the high
temperature states that are seen in simulations is shown in Fig.~\ref{distr}.
At low temperatures, there is a bowl shaped minimum
with a very broad peak at $\theta \approx \pi$. 
Thus the mean field solution predicts a value for $L_{a}$ that is
-1 at low temperatures.However, in simulations the expectation
value of $L_{a}$ is very small (almost close to zero) at low
temperatures. The way to reconcile these two statements is to note
that there are large fluctuations about the mean field solution
at low temperatures. This is apparent from the flat shape of the
effective potential at low temperatures. The second derivative of
the effective potential at the minimum is quite small and this
results in large fluctuations about the mean field solution.
This is also seen from
the distribution of $L_{a}$ at a single lattice site, which
is shown in Fig.~\ref{distr}a. This distribution has a very broad peak
at $L_{a} \approx -1$ ($\theta \approx \pi$) but there are
large fluctuations about this peak. A rough
estimate of the fluctuations about the mean field solution can
be made as follows.
The effective potential can be approximated by retaining just the
first two terms in the character expansion. This approximation is sufficient
to reproduce the form of the effective potential in Fig.~\ref{so3pot}.
The effective potential becomes
\beq
V(\theta(r))=-\sum_{r}\log(1-\cos(\theta(r)))-c\sum_{r r^{\prime}}
(1+2\cos(\theta(r)))(1+2\cos(\theta(r^{\prime})))
\quad .
\eeq
The constant $c$ is $\frac{\tilde \beta_{1}}{\tilde \beta_{0}}$.
We make the following expansion about the mean field solution:
\beq
V(\theta(r))=V(\bar \theta)+(1/2) \sum_{r r^{\prime}}\frac{\partial^{2}V}
{\partial \theta(r) \theta(r^{\prime}}|_{\bar \theta}\eta(r) \eta(r^{\prime})
+ ..
\eeq
For the $\bar \theta=\pi$ state we note that
\beq
\frac{\partial^{2}V}{\partial^{2} \theta(r) \theta(r^{\prime}}|_{\bar \theta}=0
\eeq
and
\beq
\frac{\partial^{2}V}{\partial^{2} \theta(r) }|_{\bar \theta}=(1/2) +12c
\quad .
\label{fluc}
\eeq
This leaves only the following term in the expansion
\beq
V(\theta(r))=V(\bar \theta)+(V_{1}/2)\sum_{r}\eta^{2}(r)
\eeq
where $V_{1}$ is given in Eq.~\ref{fluc}.
The partition function is given by
\beq
Z=\int d\bar \theta d\eta(r) \exp(-V(\bar \theta) \exp(-(V_{1}/2)\sum_{r}\eta^2
(r))
\eeq
The corrected value of $L_{a}$ in the presence of these fluctuations is
given by
\beq
\langle L_{a} \rangle =(1/Z)\int d\bar \theta d \eta(r) \exp(-V(\bar \theta))
\exp (\frac{-1}{2} V_{1} \sum_{r} \eta^2 (r) )
(1+2\cos(\bar \theta +\eta(r)))
\quad .
\eeq
Writing
\beq
2 \cos(\bar \theta+\eta(r))=(\exp i(\bar \theta+\eta(r))+c.c)
\eeq
and doing a gaussian integral we get the first correction to
$L_{a}$ as
\beq
<L_{a}>=1+2(-1)I \quad ,
\eeq
where $I$ is the following integral
\beq
I=\frac{\int_{-2\pi}^{2\pi}d\eta \cos(\eta)\exp(\frac{-1}{2}V_{1}\eta^{2})}
{\int_{-2\pi}^{2\pi}d\eta \exp(\frac{-1}{2}V_{1}\eta^{2})}
\quad .
\eeq
$V_{1}$ is the second derivative of the effective potential at the
minimum $\theta \approx \pi$.
These fluctuations are large at low temperatures (which is the
disordered phase)
and are small at high
temperatures ( which is the ordered phase).
Also, the above calculation is only for the leading order
correction. There will be higher order corrections which will shift the
value of $L_{a}$ from the saddle point value even further.
We have calculated this integral for some typical values in the
low temperature phase and their effect is to shift the value of $L_{a}$
(from the mean field value -1)
by a large amount. At high temperatures, the corrections are smaller as
the minima are more sharply peaked. This rough estimate of the fluctuations
shows that fluctuations about the mean field solution 
(for the $\theta \approx \pi$ minimum) increase the
value of $L_{a}$ from the mean field value.

The distribution of $L_{a}$ in the $L_{a}$ positive state is peaked at a
positive value of $L_{a}$. This can be compared with the two minima of
the effective action
at $\theta \approx 0$ and
$\theta \approx 2\pi$. Both these minima have the same value 
(close to +3) of $L_{a}$. 
In the \Laa \  negative state, the effective action has a sharp minimum at
$\theta \approx \pi$ and
this can be compared with the distribution in Fig.~\ref{distr} c, which shows
a sharp peak at $L_{a}=-1$ ($\theta \approx \pi$).
As the minima are more sharply peaked at high temperatures,
the corrections
to the mean field value will be small.
These observations show that the minima of the effective action along with
the shape of the effective action near the minima ( which represents the
effect of fluctuations about the minima) can
reproduce the structure of the high and low temperature states that are 
seen in numerical simulations.
A notable aspect of the effective action
is that the minima at $\theta \approx 0,2\pi$ are
exactly degenerate whereas the minimum at $\theta \approx \pi$ is slightly
shifted from the other two. This is not very surprising because there is
no symmetry between the $\theta \approx \pi$ and the $\theta \approx 0,2\pi$
states which requires these states to be of the same depth.
The minima at $\theta \approx 0,2 \pi$ are the same as those observed in the
$SU(2)$ theory and correspond to the deconfining phase. The minimum
at $\theta \approx \pi$ is at the same location as the minimum at low
temperatures and is only more sharply peaked.

We have also studied the effective action at a fixed value of $\beta_{a}$ 
and
varied $N_{\tau}$. Varying $N_{\tau}$ is equivalent to varying temperature
at a fixed coupling.
The purpose of this exercise is to see how the effective action
evolves with temperature in the large $\beta_{a}$ region.
This evolution is shown for two values of $\beta_{a}$, 3.5 and 5.5. The evolution
at these two couplings
is shown in Fig.~\ref{fixbet1} and Fig.~\ref{fixbet2} respectively.
At small $N_{\tau}$ (high temperatures),
there are two global minima at $\theta \approx 0$
and $\theta \approx 2\pi$, and a local minimum at $\theta \approx \pi$;
at large $N_{\tau}$ (low temperatures),
there is only one bowl shaped minimum at
$\theta \approx \pi$. This shows that as the temperature is raised at a fixed
coupling, the global minimum of the effective action shifts from 
$\theta \approx \pi$ to
$\theta \approx 0,2 \pi$.
This
again suggests that there is a finite temperature
phase transition 
at a large coupling. The two evolutions also
show that the transition to the deconfining phase takes place at
$N_{\tau}=3$ for $\beta_{a}=3.5$ and at $N_{\tau}=4$ for $\beta_{a}=5.5$.
Though the actual numbers predicted by the mean field calculation
cannot be very accurate, the analysis does serve to
demonstrate a definite trend as one increases $\beta_{a}$. As $\beta_{a}$
increases, the transition temperature becomes lower (larger $N_{\tau}$), and
at least the direction in which $\beta_{a}$ and $N_{\tau}$ are moving
is not inconsistent with general expectations.
Below we list some values of the critical coupling as a function of the
lattice size:
\begin{eqnarray}
\nonumber
N_{\tau}\ \ \ \beta_{a}^{cr} \\ \nonumber
2 \ \ \ \ 3.2 \\ \nonumber
3 \ \ \ \ 4.4 \\ \nonumber
5 \ \ \ \ 6.3 \\ \nonumber
7 \ \ \ \ 8.2 \\ \nonumber
9 \ \ \ \ 10.1
\end{eqnarray}

We now wish to point out some features of the mean field theory
which appear to be at variance
with observations in numerical simulations
The evolution of the effective potential 
(see Fig.~\ref{so3pot}) as a 
function of temperature for a fixed value of $N_{\tau}$ shows that
the minimum at $\theta \approx \pi$ continues to remain a minimum, although
a sharpened one, even for reasonably large values of $\beta_{a}$. 
Though
local minima start developing at $\theta \approx 0,2\pi$, the global minimum
still remains at $\theta \approx \pi$. It is only at a much
larger temperatures that the minima at $\theta \approx 0,2 \pi$ move below
the minimum at $\theta \approx \pi$. 
In numerical simulations, a strong metastability in the value of $L_{a}$ is observed
at high temperatures.
An ordered start always goes to the $L_{a}$ positive state wheres a random
start usually goes to the $L_{a}$ negative state. Though the mean field theory
shows that the free energy of these two states are never equal, both
states are observed in simulations depending on the initial start of the
Monte-Carlo run.
Another point is that even at large values of $\beta_{a}$,
$\theta \approx \pi$ remains a local
minimum; this may explain its appearance in simulations (with a hot start).
A cold start, which begins at $\theta \approx 0,2\pi$,
never settles to the \Laa \
negative state. It is only the hot start which
ever settles to the \Laa \  negative state.
This strong metastability in the values of the adjoint Wilson line
persists even at very high temperatures.
The other more striking feature predicted by the mean field
theory, a phase transition
from the $\theta \approx \pi$  state to the 
$\theta \approx 0,2\pi$ state, has not been directly observed in
simulations, though there are strong indications that such a phase transition
may be taking place \cite{our}.
An argument presented in \cite{our} 
showed that the deconfinement transition in the $SO(3)$ LGT would
require very large temporal lattices.
Our studies of tunnelling in \cite{our} indicated a transition 
(as a function of temperature) from 
a double peak at
$\theta \approx \pi,0,2\pi$ to a single 
peak at $\theta \approx 0, 2 \pi$. We present here one such plot of a
tunnelling study in
Fig.~\ref{tunn}. This figure shows the density of $L_{a}$
as a function
of $N_{\tau}$ on a $N_{\sigma}=7$ lattice at $\beta=3.5$. 
As $N_{\tau}$ is decreased, there is passage from the $L_{a}$ negative region
to the $L_{a}$ positive region.  This
indicates the multiple peak structure in the effective action and also the passage
from a double peak structure to a single peak structure at high
temperatures.
This should be
compared with the evolution of the effective action shown in Fig.~\ref{fixbet1}.
The comparison is only meant to show a qualitative similarity in the two, and 
finer details (such as, the location of the passage from single peak to
double peak), will certainly differ.

Next, we wish to mention a straightforward extension of the mean
field theory to the $SU(2)$ mixed action LGT.
The analysis proceeds as before and
only the coefficients of the character expansion are different in
this case. They have to be computed numerically using Eq.~\ref{chars}.
For a fixed $\beta_{a}$ and $N_{\tau}$,
the local minimum at $\theta \approx \pi$
disappears altogether for large $\beta_{f}$ ( $\beta_{f} \approx 1$),
and the effective potential has the same form
as in the $SU(2)$ LGT. This would imply that numerical simulations of the
mixed action $SU(2)$ LGT should not observe the \Laa \ 
negative state
for large values of $\beta_{f}$. This feature is also 
confirmed in numerical
simulations.

Finally, we would like to discuss some theoretical issues 
pertaining to the adjoint
Wilson line which are quite different from the fundamental Wilson line.
An appreciation of these differences is important for understanding the role
of the adjoint Wilson line in the deconfinement transition.
Firstly, the adjoint Wilson line is not an order parameter in the strict
sense and is always non-zero. Nevertheless, it can still
show critical behaviour across a phase transition. 
Another important difference between the fundamental and the adjoint Wilson line is
that the average value of the adjoint Wilson line must always be non-negative.
In the $SU(2)$ LGT,
the average value of the fundamental Wilson line is always zero in a finite system because tunnelling
between the two $Z$ related states always restores the symmetry. The
adjoint Wilson line, on the other hand, 
is not constrained to be zero by any symmetry
and is always non-zero, even on finite lattices. 
Also, the free energy interpretation in
Eq.\ref{free}
presupposes that the average value of the adjoint Wilson line
is a non-negative quantity. However, we are observing
states of negative $L_{a}$ in simulations.
Though this negative value of the adjoint Wilson line is
surprising, we note that
in the large $\beta_{a}$ region, which is the
region where we expect to make contact with the Yang-Mills
theory, the adjoint Wilson line is always positive.

From the above analysis it is evident that the mean field theory has
had some success. For the first time we are able to explain
the appearance of the
$L_{a}$ negative state and this state could not have been anticipated
apriori from any considerations. 
The structure of the high and low
temperature states observed in simulations are also explained.
The mean field theory also predicts
a phase transition 
in the large $\beta_{a}$ region.
In \cite{our}, various scenarios were suggested
to reconcile the observations made in numerical simulations of the
$SO(3)$ LGT
with theoretical
expectations. One of the scenarios suggested in \cite{our} envisioned a
phase transition from a bulk phase to a deconfining phase.
The mean field theory has provided further evidence for this transition.

The author would like to acknowledge useful discussions with Rajiv Gavai and
Saumen Datta. He would also like to thank J.~Polonyi for an
enlightening  conversation, and for suggesting to him to perform
a mean field analysis of the $SO(3)$ LGT.

\noindent
Addendum: In this paper we have not said much about the bulk, $Z(2)$ driven,
transition in the $SO(3)$ LGT, but we have concentrated more on the behaviour of
the adjoint Wilson line. In our analysis,
the effect of the bulk transition manifests itself
in the sharpening of the minimum at $\theta \approx \pi$ and the appearance
of minima at $\theta \approx 0,2\pi$
in the effective action at $\beta_{a} \approx 3$. 
In numerical simulations, the states with $L_{a}$
negative and $L_{a}$ positive are also observed immediately after the bulk
transition.

\end{section}

\newpage
\begin{figure}
% Fig1.ps 
\label{la}
\caption{The two metastable states for $L_{a}$. $L_{a}$ is plotted as a
function of Monte-Carlo sweeps/10.
The positive value is reached
after a cold start and the negative value is reached after a hot start.}
\end{figure}
\begin{figure}
% Fig2.ps 
\caption{The effective potential for the $SU(2)$ theory as a function
of $\beta_{f}$ with $N_{\tau}$ fixed to 3. The values of $\beta_{f}$ for
which the potential is shown are: 1.5,2.5,3.5,4.5,5.5,6.5. In the 
figure these correspond to parts a,b,c,d,e, and f respectively.}
\label{su2pot}
\end{figure}
\begin{figure}
% Fig3.ps
\caption{The effective potential for the $SO(3)$ theory as a function
of $\beta_{a}$ with $N_{\tau}$ fixed to 3. The values of $\beta_{a}$ for
which the potential is shown are: 1.5,2.5,3.0,3.5,4.5,5.5. In the figure
these correspond to parts a,b,c,d,e, and f respectively.}
\label{so3pot}
\end{figure}
\begin{figure}
% Fig4.ps 
\caption{The distribution of $L_{a}$ in (a) the low temperature state,
(b) the $L_{a}$ positive state, and (c) the $L_{a}$ negative state.}
\label{distr}
\end{figure}
\begin{figure}
% Fig5.ps
\caption{The effective potential for the $SO(3)$ theory as a function
of $N_{\tau}$ with a fixed $\beta_{a}$. The value of $\beta_{a}$
is 3.5 and $N_{\tau}$ takes the values: 1,2,3,4,5,7. 
In the figure these correspond to
parts a,b,c,d,e, and f respectively.}
\label{fixbet1}
\end{figure}
\begin{figure}
% Fig6.ps
\caption{The effective potential for the $SO(3)$ theory as a function
of $N_{\tau}$ with a fixed $\beta_{a}$. The value of $\beta_{a}$
is 5.5 and $N_{\tau}$ takes the values: 1,2,3,4,5,7. 
In the figure
these correspond to parts a,b,c,d,e, and f respectively.}
\label{fixbet2}
\end{figure}
\begin{figure}
% Fig7.ps
\caption{The distribution of $L_{a}$ as a function of $N_{\tau}$ at
$\beta_{a}=3.5$. The spatial lattice size was fixed at $N_{\sigma}=7$
and the temporal lattice size is indicated in the figure key.}
\label{tunn}
\end{figure}
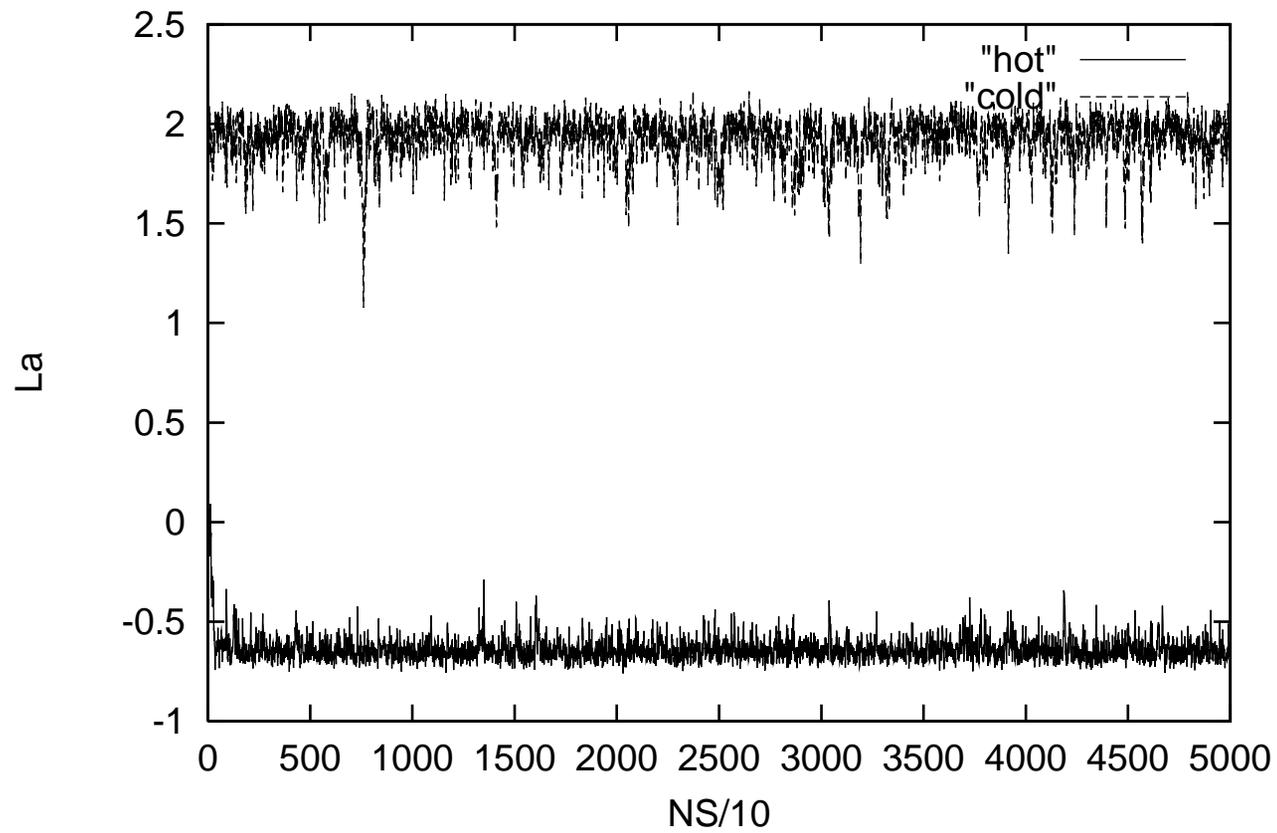
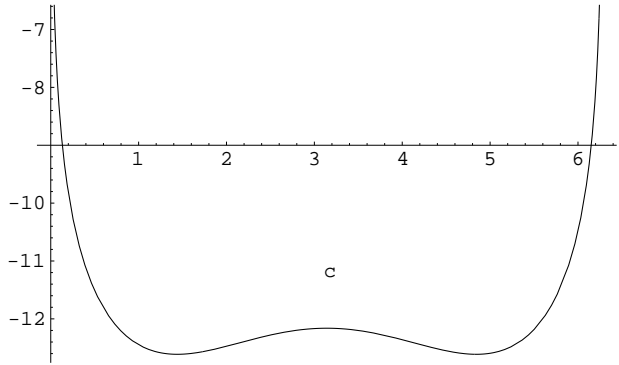
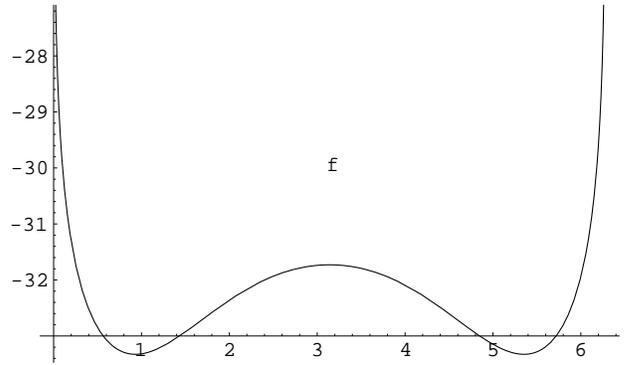
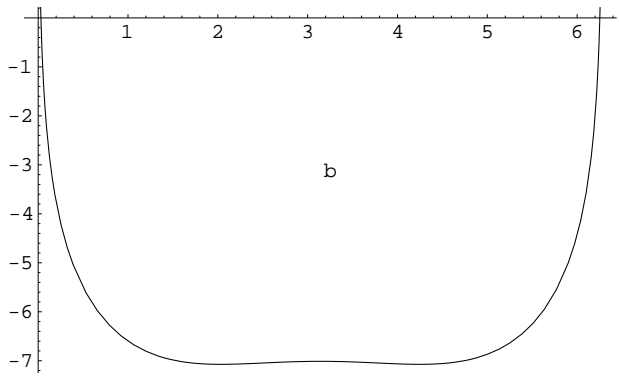
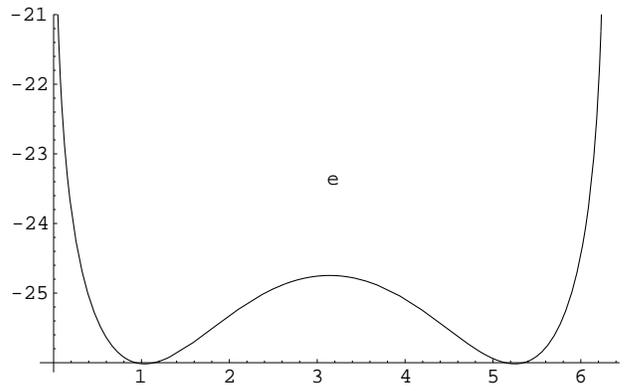
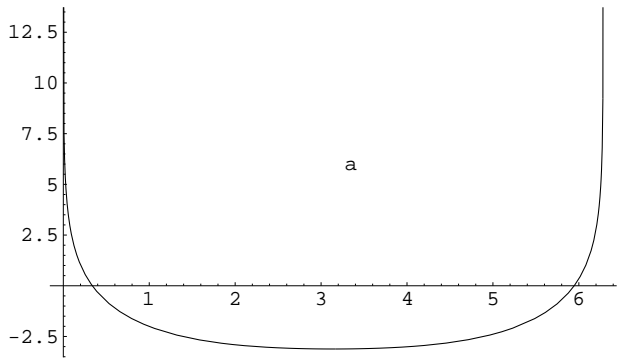
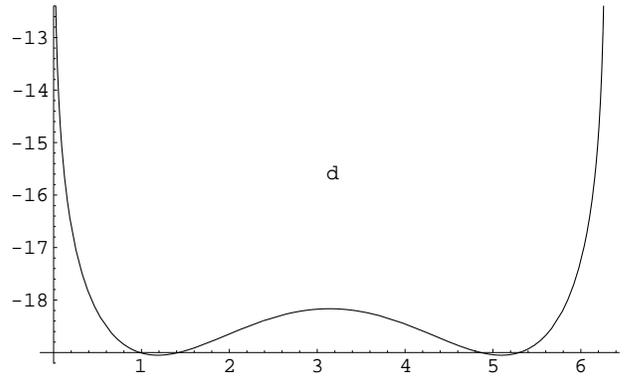
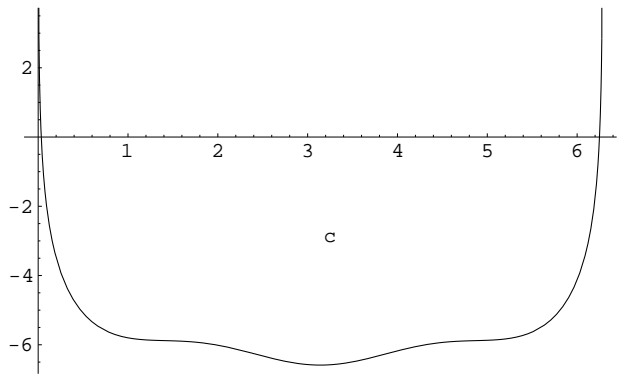
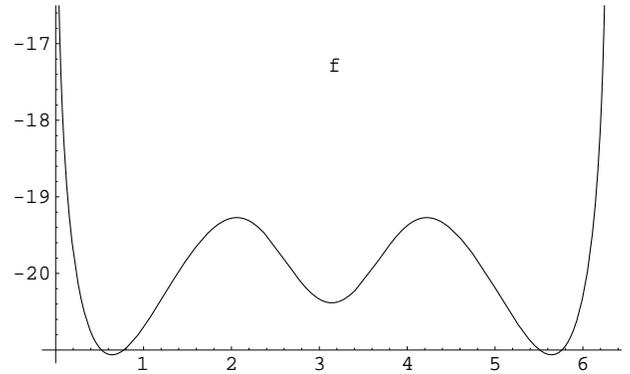
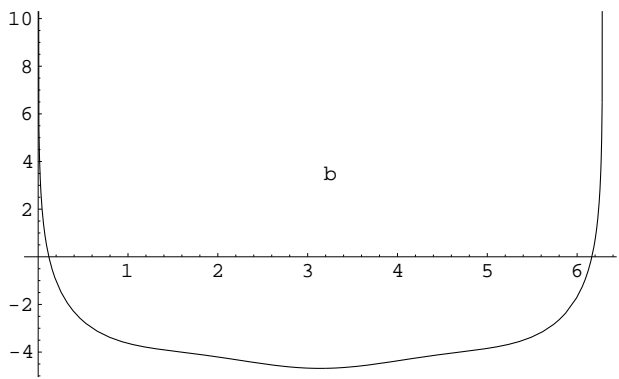
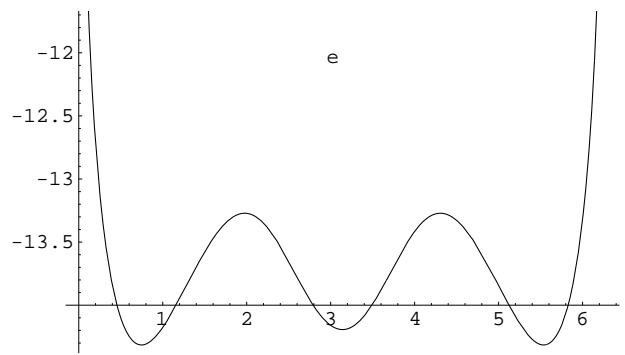
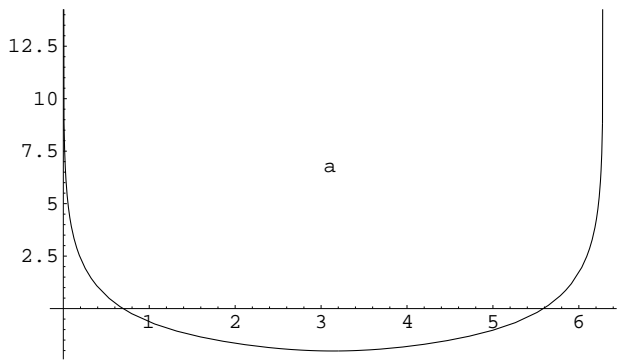
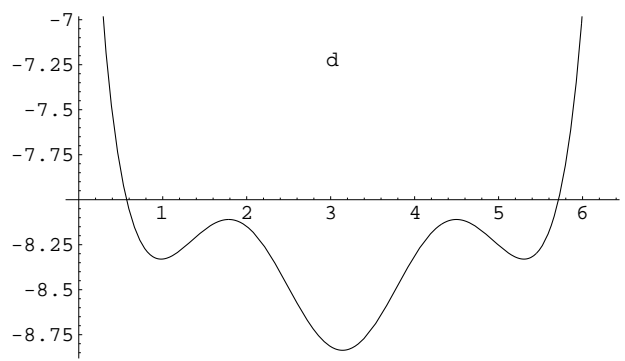
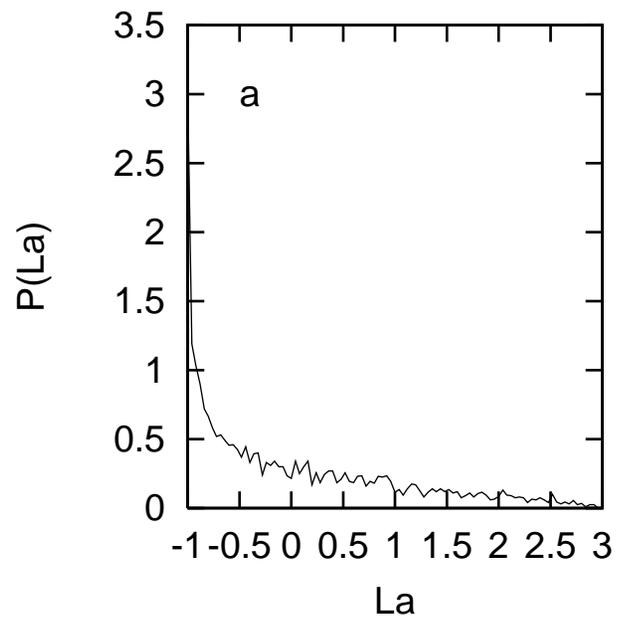
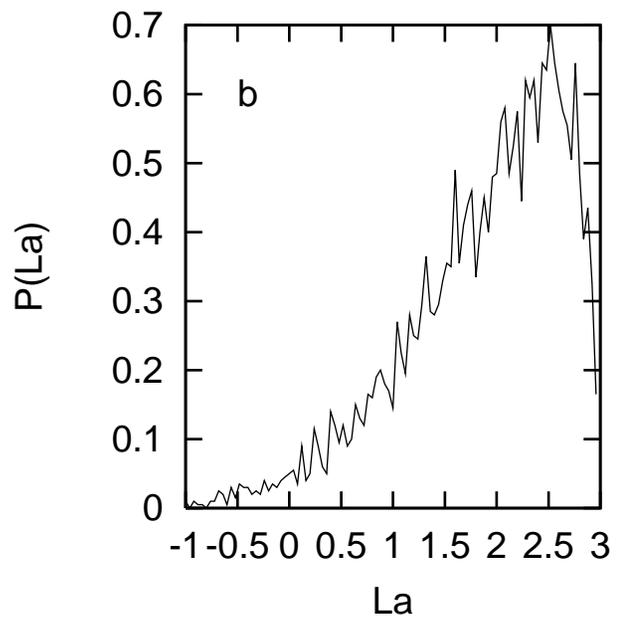
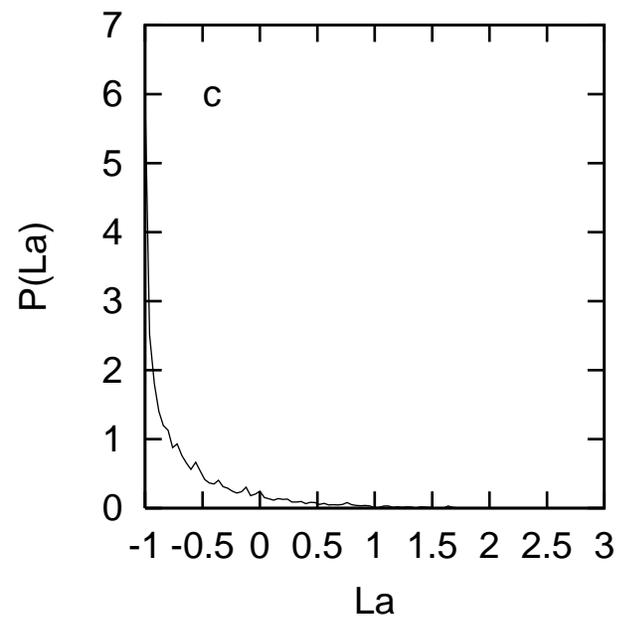
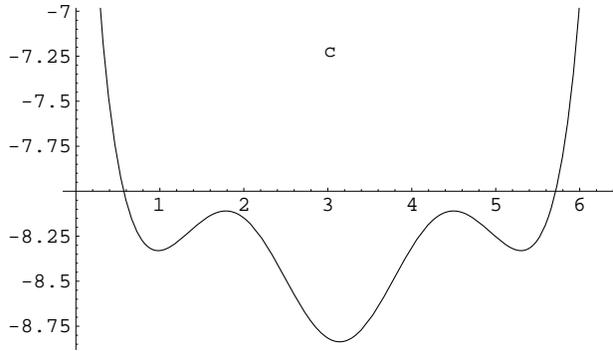
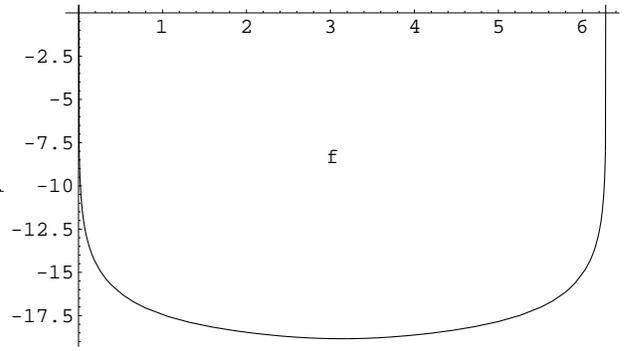
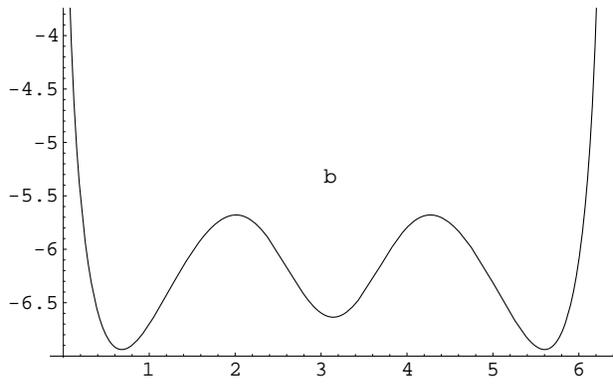
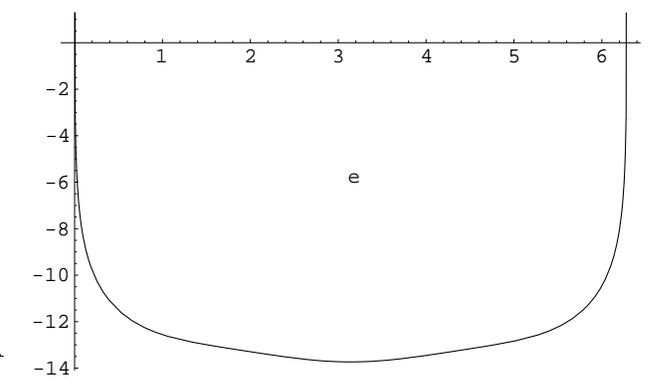
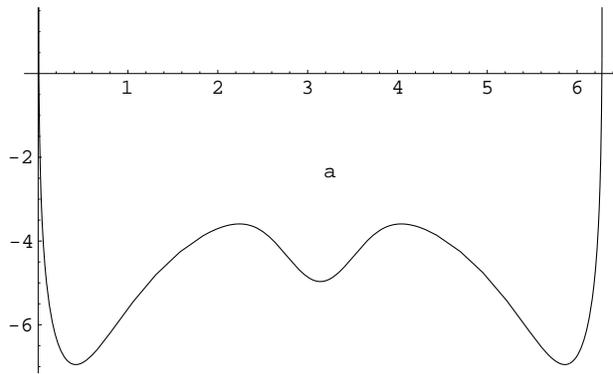
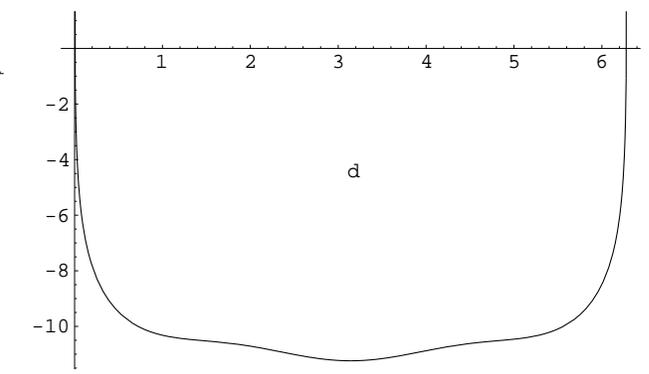
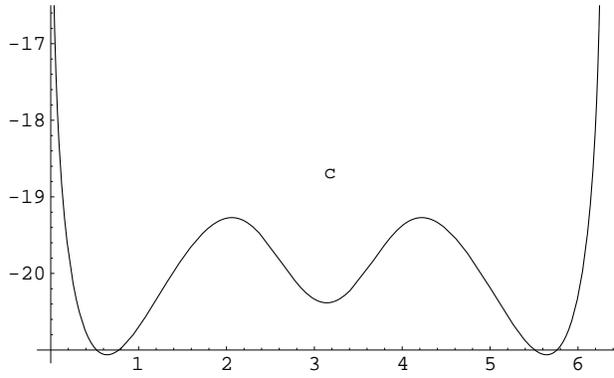
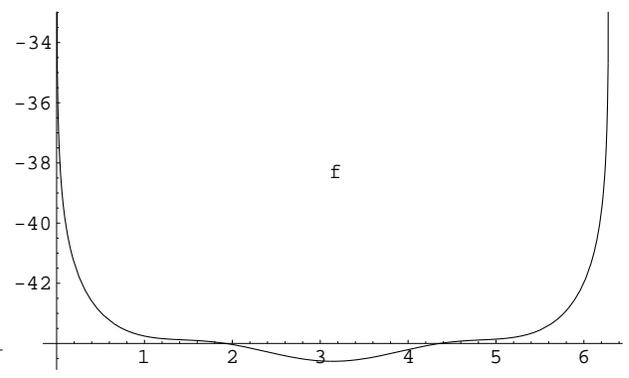
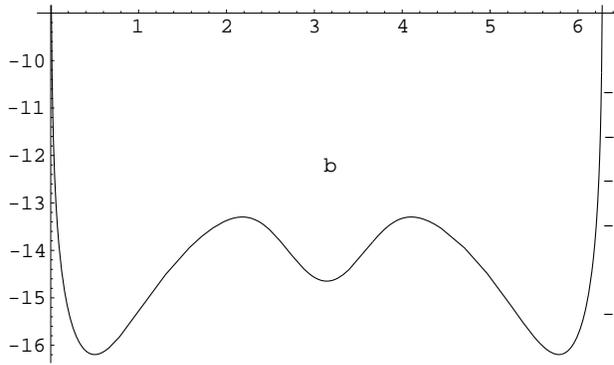
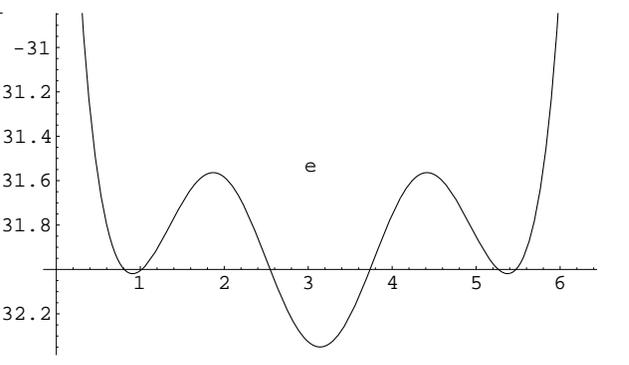
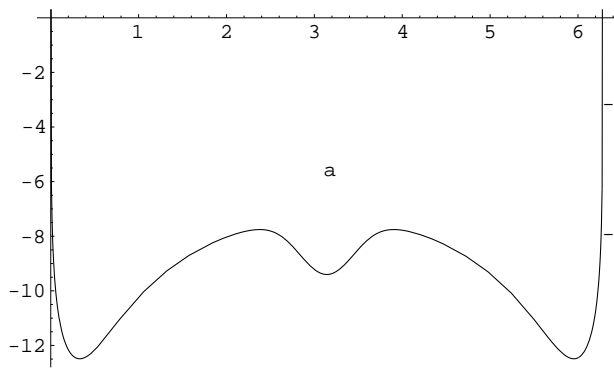
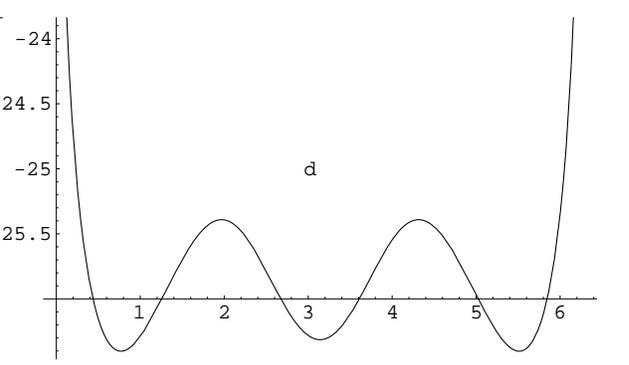
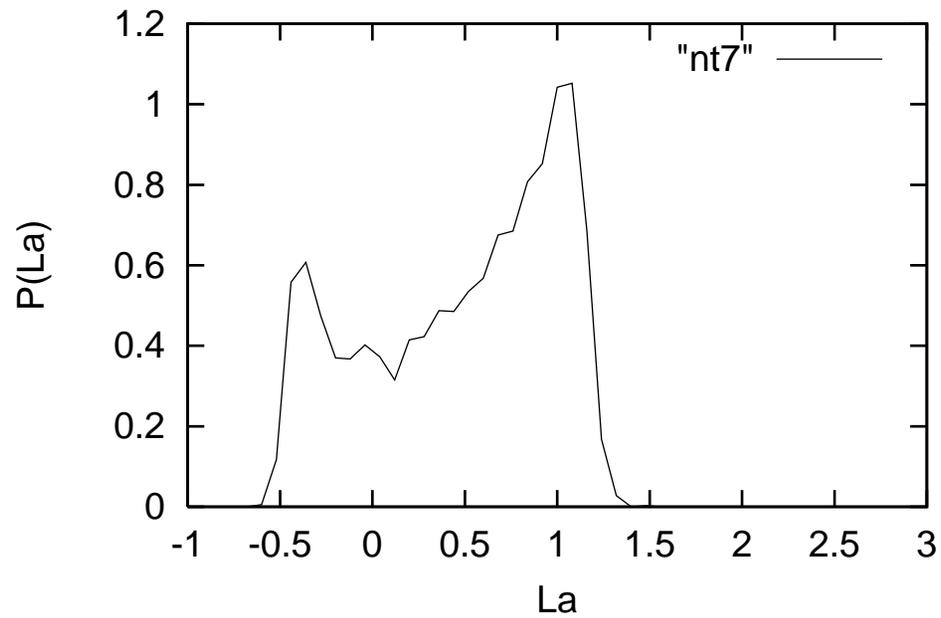
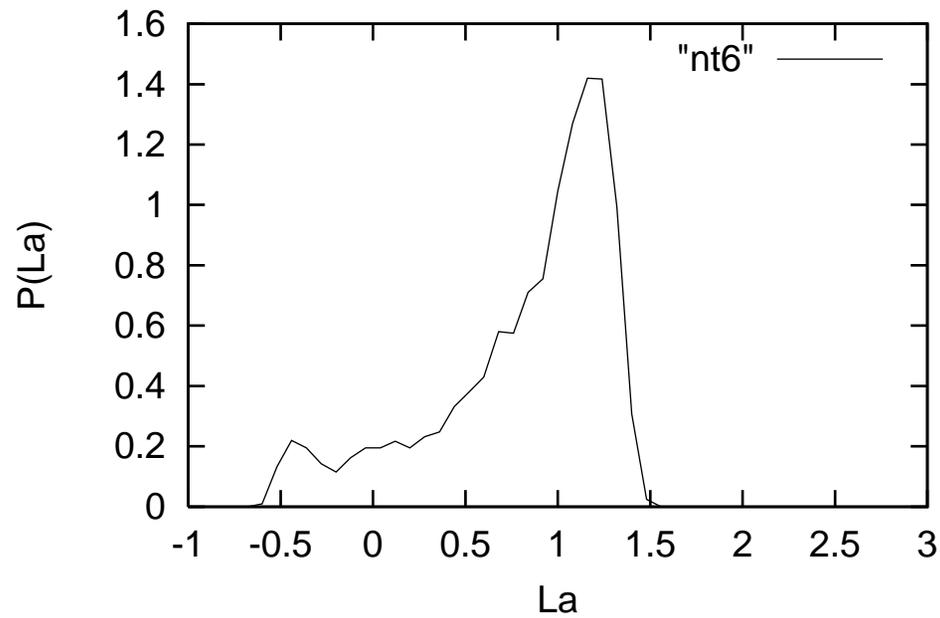
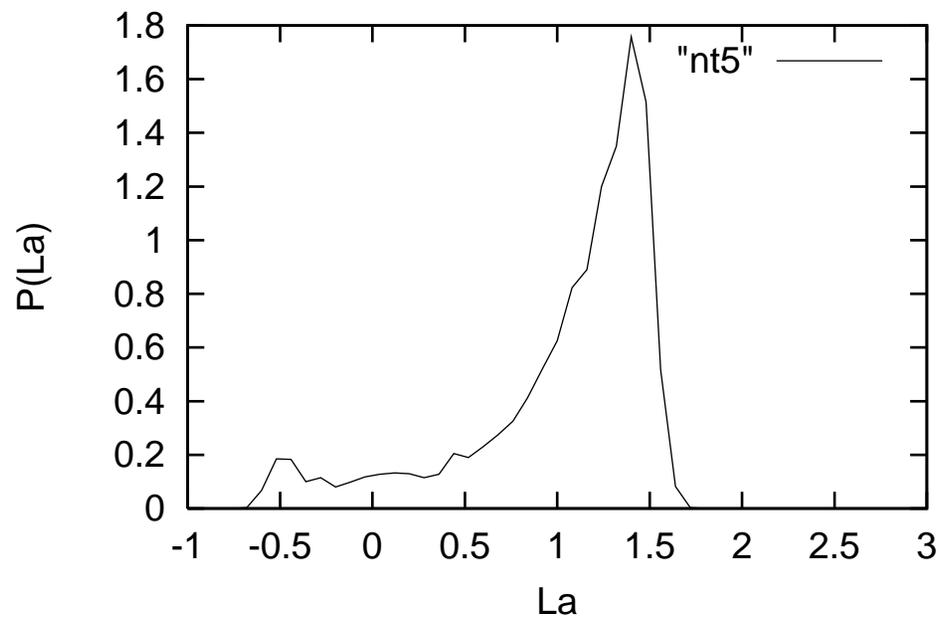
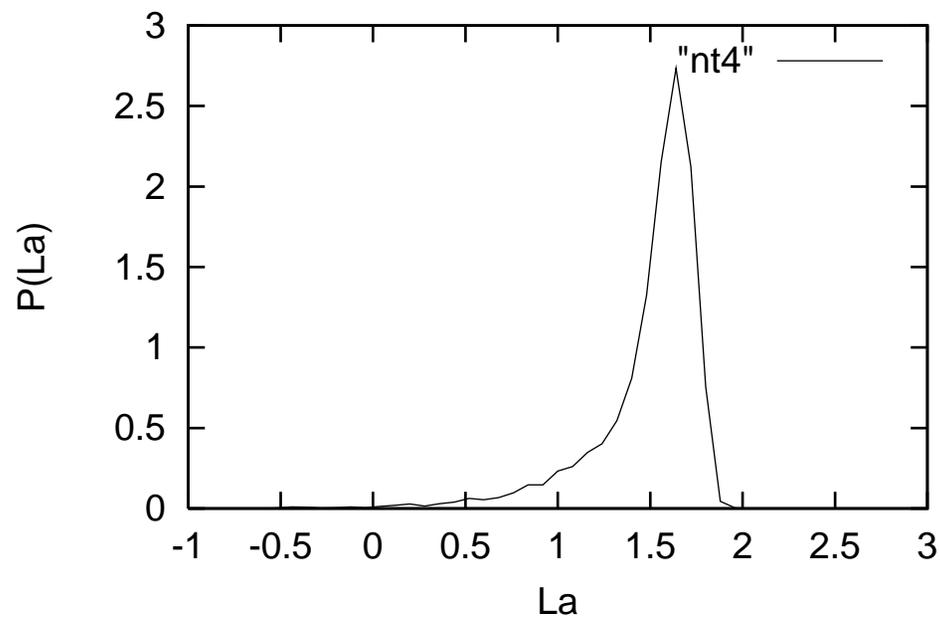
\end{document}